\documentclass[aps,prd,
letterpaper]{revtex4}
\usepackage[left=2cm, right=2cm, top=25mm, bottom=20mm]{geometry}
\usepackage{layout}
\usepackage{amsmath,amssymb,epsfig}
\usepackage{lscape}
\usepackage{color}
\usepackage{graphicx}
\usepackage{url}
\usepackage{comment}
\usepackage{lineno}

\newcommand\etal{{\em et al.}}
\newcommand\dbd{\ensuremath{0\nu\beta\beta}}

\begin{document}

\title{{\bf {\huge CUPID}} \\ CUORE (Cryogenic Underground Observatory for Rare Events) \\ Upgrade with Particle IDentification}




\author{G.~Wang}
\author{C.L.~Chang}
\author{V.~Yefremenko}
\affiliation{High Energy Physics Division, Argonne National Laboratory, Argonne, IL, USA}

\author{J.~Ding}
\author{V.~Novosad}
\affiliation{Materials Science Division, Argonne National Laboratory, Argonne, IL, USA}

\author{C.~Bucci}
\author{L.~Canonica}
\author{P.~Gorla}
\author{S.S.~Nagorny}
\author{C.~Pagliarone}
\altaffiliation{Also with: University of Cassino,  Cassino Frosinone, Italy}
\author{L.~Pattavina}
\author{S.~Pirro}
\author{K.~Schaeffner}
\affiliation{INFN - Laboratori Nazionali del Gran Sasso, Assergi (AQ), Italy}

\author{J.~Feintzeig}
\author{B.K.~Fujikawa}
\author{Y.~Mei}
\affiliation{Nuclear Science Division, Lawrence Berkeley National Laboratory, Berkeley, CA, USA}

\author{E.B.~Norman}
\author{B.S.~Wang}
\affiliation{Department of Nuclear Engineering, University of California, Berkeley, CA, USA}

\author{T.I.~Banks}
\author{Yu.G.~Kolomensky}
\altaffiliation{Also with: Physics Division, Lawrence Berkeley National Laboratory, Berkeley, CA, USA}
\author{R.~Hennings-Yeomans}
\author{T.M.~O'Donnell}
\author{V.~Singh}
\affiliation{Department of Physics, University of California, Berkeley, USA}

\author{N.~Moggi}
\author{S.~Zucchelli}
\affiliation{Universit\`a di Bologna and INFN Bologna, Bologna, Italy}

\author{L.~Gladstone}
\author{L.~Winslow}
\affiliation{Massachusetts Institute of Technology, Cambridge, MA, USA}

\author{D.R.~Artusa, F.T.~Avignone III, R.J.~Creswick, H.A.~Farach, C.~Rosenfeld, J.~Wilson}
\affiliation{Department of Physics and Astronomy, University of South Carolina, Columbia, SC, USA}

\author{J.~Lanfranchi, S.~Sch\"onert, M.~Willers}
\affiliation{Technische Universit\"at M\"unchen, Physik-Department E15, Garching, Germany}

\author{S.~Di Domizio, M.~Pallavicini}
\affiliation{Dipartimento di Fisica, Universit\`a di Genova and INFN - Sezione di Genova, Genova, Italy}

\author{M.~Calvo, A.~Monfardini}
\affiliation{Institut N\'eel,CNRS and Universit\'e de Grenoble, INP, Grenoble,
France}

\author{C.~Enss, A.~Fleischmann, L.~Gastaldo}
\affiliation{Kirchhoff-Institute for Physics, University of Heidelberg,
Heidelberg, Germany} 

\author{R.S.~Boiko, F.A.~Danevich, V.V.~Kobychev}
\author{D.V.~Poda}
\altaffiliation{Also with: Centre de Sciences Nucleaires et de
  Sciences de la Matiere (CSNSM), CNRS/IN2P3, Orsay, France}
\author{O.G.~Polischuk}
\altaffiliation{Also with: INFN, Sezione di Roma ``La Sapienza'', Rome, Italy}
\author{V.I.~Tretyak}
\altaffiliation{Also with: INFN, Sezione di Roma ``La Sapienza'', Rome, Italy}
\affiliation{Institute for Nuclear Research, Kyiv, Ukraine}

\author{G.~Keppel, V.~Palmieri}
\affiliation{INFN - Laboratori Nazionali di Legnaro, Legnaro, Italy}

\author{K.~Kazkaz, S.~Sangiorgio, N.~Scielzo}
\affiliation{Lawrence Livermore National Laboratory, Livermore, CA, USA}

\author{K.~Hickerson, H.~Huang}
\affiliation{Department of Physics and Astronomy, University of California, Los Angeles, CA, USA}

\author{M.~Biassoni, C.~Brofferio, S.~Capelli, D.~Chiesa, M.~Clemenza,
O.~Cremonesi, M.~Faverzani, E.~Ferri, E.~Fiorini, A.~Giachero,
L.~Gironi, C.~Gotti, A.~Nucciotti, M.~Pavan, G.~Pessina,
E.~Previtali, C.~Rusconi, M.~Sisti, F.~Terranova}
\affiliation{INFN sez.~di Milano Bicocca and 
Dipartimento di Fisica, Universit\`a di Milano Bicocca, Milano, Italy}

\author{A.S.~Barabash, S.I.~Konovalov, V.V.~Nogovizin, V.I.~Yumatov}
\affiliation{State Scientific Center of the Russian Federation - Institute of Theoretical and Experimental Physics (ITEP), Moscow, Russia}

\author{F.~Petricca, F.~Pr\"obst, W.~Seidel}
\affiliation{Max-Planck-Institut f\"ur Physik, D-80805 M\"unchen, Germany}

\author{K.~Han, K.M.~Heeger, R.~Maruyama, K.~Lim}
\affiliation{Wright Laboratory, Department of Physics, Yale University, New Haven, CT, USA}

\author{N.V.~Ivannikova, P.V.~Kasimkin, E.P.~Makarov, V.A.~Moskovskih,
V.N.~Shlegel, Ya.V.~Vasiliev, V.N.~Zdankov}
\affiliation{Nikolaev Institute of Inorganic Chemistry, SB RAS, Novosibirsk, Russia}

\author{A.E.~Kokh, V.S.~Shevchenko, T.B.~Bekker}
\affiliation{Sobolev Institute of Geology and Mineralogy, SB RAS, Novosibirsk, Russia}

\author{A.~Giuliani, P.~de Marcillac, S.~Marnieros, E.~Olivieri}
\affiliation{Centre de Sciences Nucl\`eaires et de Sciences de la Mati\`ere (CSNSM), CNRS/IN2P3, Orsay, France}

\author{L.~Taffarello}
\affiliation{INFN - Sezione di Padova, Padova, Italy}

\author{M.~Velazquez}
\affiliation{Institut de Chimie de la Mati\`ere Condens\'e de Bordeaux (ICMCB), CNRS, 87, Pessac, France}

\author{F.~Bellini}
\author{L.~Cardani}
\altaffiliation{Also with: Physics Department, Princeton University, Princeton, NJ, USA}
\author{N.~Casali, I.~Colantoni, C.~Cosmelli, A.~Cruciani, I.~Dafinei, F.~Ferroni, S.~Morganti, P.J.~Mosteiro, F.~Orio, C.~Tomei, V.~Pettinacci, M.Vignati}
\affiliation{Dipartimento di Fisica, Universit\`a di Roma ``La Sapienza'' and INFN - Sezione di Roma, Roma, Italy}

\author{M.G.~Castellano}
\affiliation{IFN-CNR, Via Cineto Romano, Roma, Italy}

\author{C.~Nones}
\affiliation{Service de Physique des Particules, DSM/IRFU, CEA-Saclay, Saclay, France}

\author{T.D.~Gutierrez}
\affiliation{Physics Department, California Polytechnic State University, San Luis Obispo, CA, USA}

\author{X.G.~Cao, D.Q.~Fang, Y.G.~Ma, H.W.~Wang, X.G.~Deng}
\affiliation{Shanghai Institute of Applied Physics (SINAP), Shanghai, China}

\author{A.~Cazes, M.~De Jesus}
\affiliation{Institut de Physique Nucl\`eaire de Lyon, Universit\'e Claude Bernard, Lyon 1, Villeurbanne, France}

\author{B.~Margesin}
\affiliation{Fondazione Bruno Kessler, Trento, Italy }

\author{E.~Garcia}
\author{M.~Martinez}
\altaffiliation{Also with: Dipartimento di Fisica, Universit\`a di Roma ``La Sapienza'' and INFN - Sezione di Roma, Roma, Italy}
\author{J.~Puimedon, M.L.~Sarsa}
\affiliation{Universidad de Zaragoza, Laboratorio de Fisica Nuclear y Astroparticulas, Zaragoza, Spain}

\date{\today}

\begin{abstract} 
CUPID is a proposed future tonne-scale bolometric
neutrinoless double beta decay (\dbd) experiment 
to probe the Majorana nature of neutrinos and discover Lepton Number
Violation in the so-called inverted hierarchy region of the neutrino
mass. 
CUPID will be built on experience, expertise and
lessons learned in CUORE, and will exploit the current CUORE
infrastructure as much as possible. In order to achieve its ambitious
science goals, CUPID aims to increase the source mass and
dramatically reduce the backgrounds in the region of interest.
This requires isotopic enrichment, upgraded
purification and crystallization procedures, new detector
technologies, a stricter material selection, and possibly new shielding
concepts with respect to the state of the art deployed in CUORE. 
This document reviews the science goals of CUPID, defines the
scope for the near-term R\&D activities, and presents a
roadmap towards mounting this next-generation experiment. A separate
document discusses the extensive R\&D program in more detail. 
\end{abstract}

\maketitle

\section{Introduction}\label{sec:intro}

The observation of neutrinoless double-beta decay ($0\nu\beta\beta$)
would unambiguously establish the Lepton Number violation, and
indicate that neutrinos are Majorana particles, 
i.e. they are their own anti-particles.  The rate of the process is
sensitive to the effective Majorana neutrino mass. Determining whether
neutrinos are Majorana or Dirac particles and measuring their masses
are among the highest priorities in neutrino physics. The answer will
have important 
implications for astrophysics, cosmology, as well as particle
physics. Addressing this question 
has become an even higher priority since the recent apparent discovery
of the long-sought Higgs boson.  A Majorana neutrino mass is not
generated by the Higgs mechanism and Majorana particles are not
accommodated in the Standard Model. Thus, discovery of the Majorana
nature of neutrinos would provide a clear indication of new physics
beyond the Standard Model.

CUORE~\cite{CUORE}, the Cryogenic Underground Observatory for Rare
Events, promises to be one of the most sensitive $0\nu\beta\beta$
experiments this decade. Using a bolometric array of 988 750-g
crystals of natural TeO$_2$, it will begin to explore neutrino
mass values in the so-called inverted mass hierarchy. CUORE is an
established project within the Italian (INFN) and US (DOE and NSF)
funding agencies. The detector is currently under construction at the
Laboratori Nazionali del Gran Sasso (LNGS) in Assergi, Italy, and is
expected to start operations within a year.

The purpose of this document, drafted by the CUPID Steering
Committee\footnote{CUPID steering committee: F.T.~Avignone,
F.~Bellini, C.~Bucci, O.~Cremonesi, F.~Ferroni, A.~Giuliani, P.~Gorla,
K.M.~Heeger, Yu.G.~Kolomensky, M.~Pallavicini, M.~Pavan, S.~Pirro,
M.~Vignati}, is to define a possible follow-up to the present
CUORE~\cite{CUORE} experiment, after CUORE completes its scientific
mission in about 5 years of data taking. The
next experiment will be based on experience, expertise, and
lessons learned in CUORE; thus we refer to this future project in the
following as CUORE Upgrade with Particle ID (CUPID)~\cite{CUPID}. We will first
discuss the scientific objective of CUPID; we will then list 
the important near-term R\&D activities 
which aim to develop
technologies capable of achieving the desired science goal; we will
finally indicate a time schedule for CUPID definition,
anticipating that the general goal is to select the CUPID
technology in the next few years, so that a Conceptual Design Report
(CDR) can be produced.

Continued progress towards CUORE operations motivates planning of a
next-generation double beta decay experiment with bolometric
detectors.  CUORE-0, the first CUORE tower which has been operating in
the Cuoricino cryostat for nearly two years, shows excellent
performance in terms of background and detector resolution~\cite{Q0};
all CUORE towers have been fully built; the CUORE cryostat has reached
its design base temperature~\cite{ColdestCubicMeter} and is under
final stages of commissioning~\cite{CUORE}. The time has come to plan
a future use, beyond CUORE, of the existing CUORE facilities with
improved detectors aiming at an even higher sensitivity to
\dbd. In fact, an
upgrade of the present technology or a development of a new one
requires sufficient head start in order to be ready in time by the end
of the present CUORE program.

\section{CUORE}\label{sec:cuore}

CUORE will be one of the most sensitive \dbd\
experiments of this decade. Using a bolometric array of 988 750~g
crystals of natural TeO$_2$, it will begin to explore neutrino
mass values in the inverted mass hierarchy. 
CUORE is in the final phase of construction at LNGS and is expected to
start operations within a year. Construction of all 19 detector towers
is now complete. The cryogenic system has been
completely assembled and commissioning is steadily progressing. 

With an expected background of 10~counts/(keV ton year) and an energy
resolution of 5 keV FWHM in the $0 \nu \beta \beta$ region of
interest (ROI), CUORE is projected to reach a 90\% C.L. sensitivity of
$T_{1/2} > 1 \times 10^{26}$~y after five years of
operation~\cite{Sensitivity}, which corresponds to a range of the
effective Majorana neutrino masses of 
$\langle m_{\beta \beta} \rangle < 51-133$~meV, depending on the
estimate of the nuclear matrix element. 

The CUORE concept of a bolometric $0\nu\beta\beta$ detector has
already been successfully demonstrated through the operation of  two
medium size prototypes: Cuoricino and CUORE-0. The latter in
particular has been built strictly following the same protocols used
for the construction of the CUORE  detector.   
CUORE-0 has demonstrated the viability of the 
key performance parameters: the energy resolution of the detectors,
and the background level in the region dominated by surface
contamination from $^{238}$U and $^{232}$Th. However, two of the most
challenging aspects still need to be demonstrated through the
successful operation of CUORE: long-term operation in stable
conditions of a ton-sized 
bolometric detector, and validation of the background model~\cite{CUPID,CuoreBB}. 

The CUORE cryostat and dilution refrigerator represent a breakthrough
in the currently available technology~\cite{ColdestCubicMeter} and
their successful operation 
will be a significant milestone for development of bolometric
experiments.  Scientific success of CUORE is a required condition for
future developments.  Based on careful material assays and a set of
dedicated measurements, the CUORE background budget is in the range of
the design value of 10~counts/(keV ton year). The background model,
based on Cuoricino and CUORE-0, has limitations due to the relatively
large $\gamma$ background from the Cuoricino cryostat and limited
exposure. In particular,  precise evaluation of the relative 
contributions of the $\alpha$ and $\beta/\gamma$ components in the
$^{130}$Te ROI is only possible with a 
large-scale detector like CUORE. In this
respect, CUORE itself can be considered  a very important R\&D effort
for the development of a next-generation bolometric \dbd\
experiment.

\section{Scientific Objective}\label{sec:science}

CUPID  is a proposed bolometric $0 \nu \beta \beta$ experiment
which aims at a sensitivity to the effective Majorana neutrino mass on
the order of 10 meV, covering entirely the so-called
inverted hierarchy region of the neutrino mass pattern. 
CUPID will be designed in such a way that, if the neutrino
is a Majorana particle with an effective mass in or above the
inverted hierarchy region ($\sim 15-50$~meV), then
CUPID will observe $0 \nu \beta \beta$ with a sufficiently high
confidence (significance of at least $3 \sigma$). This level of
sensitivity corresponds to a \dbd\ lifetime of
$10^{27} - 10^{28}$ years, depending on the isotope. This primary
objective poses a set of technical challenges: the sensitive detector
mass must be in the range of several hundred kg to a ton of the
isotope, and the background must be close to zero at the
ton~$\times$~year exposure scale in the ROI of a few keV around \dbd\
transition energy.

The scientific goals outlined above can be achieved
with a bolometric experiment like CUORE at LNGS, with cost-effective
upgrades. The required improvements would include: (a) isotopic
enrichment of the element of choice; (b) active rejection of alpha and
surface backgrounds in detector materials; (c) further reduction
(compared to CUORE) in the gamma backgrounds by careful material and
isotope selection and active veto of multi-site events; (d)
improvements in energy resolution; and (e) further reduction in
cosmogenically generated radioactive backgrounds. 

Successful cryogenic operations of CUORE, as well as the CUORE
experience in ultra-clean assembly of bolometers and a cryostat
system, are critical for demonstrating the viability of a future
experiment. Groups in both Europe and
the US are engaged in an active R\&D program~\cite{CUPID-RandD} 
along all the directions 
outlined above. With the current support from the US and European
funding agencies, we are investigating the cost and purity of
bolometric crystals highly enriched in $^{130}$Te as well as other
potential isotopes ($^{82}$Se, $^{100}$Mo, and $^{116}$Cd), studying
background rejection with scintillation, Cherenkov radiation,
ionization, and pulse-shape discrimination, and testing novel materials
and sensor technologies. Experience with the CUORE assembly will allow
us to further refine and optimize the process of putting together a
1-ton scale detector. Once CUORE is operational we expect the vigor
of these R\&D activities to ramp up, with the goal of preparing a full
proposal for an upgraded bolometric experiment with
$\mathcal{O}(10~\mathrm{meV})$  Majorana
mass sensitivity. The goals of the CUPID program, depending on the
ultimate isotope of choice, are listed in Table~\ref{tab:sens}.
\begin{table}[tb]
\caption{CUPID sensitivity goals}
\label{tab:sens}
\begin{center}
\begin{tabular}{l|ll}
\hline
\hline
Parameter & \multicolumn{2}{l}{Projected value and/or range} \\
\hline
\hline
Readiness for construction & \multicolumn{2}{l}{2018 (technical limit)} \\
\hline
Construction time & \multicolumn{2}{l}{5 years} \\
\hline
Total fiducial mass (kg) & TeO$_2$ & 750 \\
                         & ZnMoO$_4$ & 540 \\
                         & ZnSe & 670 \\
                         & CdWO$_4$ & 980 \\
\hline
Isotope fiducial mass (kg) & $^{130}$Te & 543 \\
                           & $^{100}$Mo & 212 \\
			   & $^{82}$Se & 335 \\
			   & $^{116}$Cd & 283 \\
\hline
Energy resolution at endpoint (FWHM) & \multicolumn{2}{l}{$<5$~keV} \\
\hline
Event selection efficiency in fiducial volume & \multicolumn{2}{l}{75-90\%} \\
\hline
Background within FWHM of endpoint
& \multicolumn{2}{l}{$<0.02$ counts/(ton$\cdot$year)} \\
\hline
90\% C.L. \dbd\ lifetime limit for 10 year run ($10^{27}$ years) &
  $^{130}$Te & $5.1$ \\
& $^{100}$Mo & $2.2$ \\
& $^{82}$Se  & $4.2$ \\
& $^{116}$Cd & $3.0$  \\
\hline
90\% C.L. $m_{\beta\beta}$ 
limit for 10 year run (90\% C.L.) (meV) &
  $^{130}$Te & 6--15 \\
& $^{100}$Mo & 6--17 \\
& $^{82}$Se  & 6--19 \\
& $^{116}$Cd & 8--15  \\
\hline
\dbd\ lifetime discovery sensitivity ($3\sigma$) in 10 years &
  $^{130}$Te & $4.9$ \\
& $^{100}$Mo & $2.1$ \\
& $^{82}$Se  & $4.0$ \\
& $^{116}$Cd & $2.9$  \\
\hline
$m_{\beta\beta}$  discovery sensitivity ($3\sigma$) in 10 years &
  $^{130}$Te & 6--15 \\
& $^{100}$Mo & 7--17 \\
& $^{82}$Se  & 6--19 \\
& $^{116}$Cd & 8--15  \\
\hline\hline
\end{tabular}
\end{center}
\end{table}


A detailed description of the ongoing R\&D efforts is given in a
separate document~\cite{CUPID-RandD}. Below, we present the
roadmap for converging on an experimental proposal in the next few
years.

\section{Key Goals for the CUPID Technology} \label{sec:selec}

In the following, we discuss
the very general guidelines that will drive the CUPID detector technology and
the choice of isotope.
The CUPID collaboration will set up a formal process in order to make
technical choices  involving the isotope, the
detector technology, and additional measures required to achieve the
sensitivity goals of the new experiment.

\begin{itemize}
\item The primary goal of CUPID is sensitivity of 10--15~meV to the
effective neutrino mass. This requires a detector with an active
isotope mass of order ton and a background level of
$\lesssim 10^{-1}$~counts/(ton$\cdot$y) in the region of interest. While
this background level cannot be verified directly before CUPID is in
operation, the chosen technology should prove convincingly that this
target can be achieved, by means of dedicated experimental tests and
verifiable simulations. Key performance parameters such as internal
radioactivity of the crystals and other detector materials, alpha/beta
and/or surface/bulk event rejection capability, alpha backgrounds
above $2.6$~MeV, sensitivity to cosmogenic backgrounds, energy
resolution, and others will be taken into account. 
\item One of the key features of the bolometric technology is the
excellent energy resolution. It is important that this feature is
maintained in the CUPID approach. Ideally, the CUPID energy resolution
should not be worse than that achieved by CUORE. This must be proven
in dedicated experimental tests with crystals of the size to be used
for CUPID.   
\item A tonne-scale bolometric detector will imply $\mathcal{O}(1000)$
single detectors. An appealing feature of the bolometric technology is
its scalability: from single-module devices, to a full-tower
demonstrator prototypes, to the full CUORE-sized array. The
next-generation technology should maintain this feature. 
\item A chosen technology must demonstrate reproducibility
in terms of technical performance (energy resolution, pulse shape,
noise features). The detector behavior should  therefore be tested
with an array of at least 8 modules and, if possible, larger, operated
underground under conditions as similar as possible to those expected
in the CUPID experiment in terms of base temperature, vibration
level, read-out, and electronics configuration. Detector assembly
reproducibility and radiopurity similar to or better than that achieved
in CUORE must be feasible. 
\item The cost and schedule of the enrichment process and of the
crystal production must be compatible with a timely realization of the
experiment. This compatibility must be proven by means of already
established contacts with the companies or institutions responsible
for enrichment and crystal production, which will be invited to
provide preliminary but realistic cost figures and production
timescales.
\item CUPID technology should be as compatible as possible with the existing CUORE infrastructure, in terms of mechanical coupling, cryogenics, readout, and DAQ features. 
\end{itemize}

In the next two years, the R\&D efforts~\cite{CUPID-RandD} will
proceed vigorously. 
CUORE operations and background measurements will inform the 
decision about the detector technology (as well as the isotope choice).
We foresee producing the CDR and forming the international collaboration on a similar timescale. 

\section{Conclusions}

Tonne-scale bolometric detectors with background rejection
capabilities beyond that of CUORE have the
potential to convincingly discover the Majorana nature of
neutrinos in the so-called Inverted Neutrino Mass
Hierarchy~\cite{CUPID}.
Following commissioning of the CUORE detector and a brief period of
vigorous R\&D activities aiming to demonstrate that the science goals
of CUPID can be effectively achieved, we will 
complete the design of a future \dbd\ experiment based on the CUORE
experience and -- to the largest 
possible extent -- on the CUORE infrastructure. The objective of this
future experiment is to discover \dbd\ decay and, therefore, establish
violation of Lepton Number if the neutrino is a Majorana
particle with the effective mass in or above the inverted hierarchy
range. 



\end{document}